\documentclass[a4paper,reqno,12pt]{amsart}
\usepackage{amsmath,amsfonts,amssymb,amsthm,amstext,eucal}
\usepackage{bbold}
\usepackage{array}
\usepackage{eepic}
\usepackage[T1]{fontenc}
\usepackage[latin1]{inputenc}
\DeclareMathOperator{\AdS}{AdS}

\DeclareMathOperator{\tr}{tr}

\renewcommand{\d}{\partial}
\newcommand{\1}{\mathbb{1}}

\newcommand{\ZZ}{\mathbb{Z}}
\newcommand{\RR}{\mathbb{R}}

\newcommand{\ISO}{\mathrm{ISO}}
\newcommand{\SO}{\mathrm{SO}}

\newcommand{\fso}{\mathfrak{so}}

\newcommand{\fs}{\mathfrak{s}}

\newcommand{\fS}{\mathfrak{S}}
\newcommand{\fg}{\mathfrak{g}}

\newcommand{\fp}{\mathfrak{p}}
\newcommand{\fk}{\mathfrak{k}}
\newcommand{\half}{\tfrac{1}{2}}

\begin{document}
\title[Homogeneous Fluxes, Branes and Maximal Supersymmetry]%
{Homogeneous fluxes, branes and a maximally supersymmetric solution of
  M-theory}
\author[Figueroa-O'Farrill]{José Figueroa-O'Farrill}
\address{\begin{flushright}
    Department of Mathematics and Statistics\\
    The University of Edinburgh\\
    James Clerk Maxwell Building\\
    King's Buildings, Mayfield Road\\
    Edinburgh EH9 3JZ\\
    Scotland\\
  \end{flushright}}
\email{j.m.figueroa@ed.ac.uk}
\author[Papadopoulos]{George Papadopoulos}
\address{\begin{flushright}
    Department of Mathematics\\
    King's College\\
    University of London\\
    Strand, London WC2R 2LS\\
    England\\
  \end{flushright}}
\email{gpapas@mth.kcl.ac.uk}
%\date{\today}
\begin{abstract}
  We find M-theory solutions with homogeneous fluxes for which the
  spacetime is a lorentzian symmetric space.  We show that generic
  solutions preserve sixteen supersymmetries and that there are two
  special points in their moduli space of parameters which preserve
  all thirty-two supersymmetries.  We calculate the symmetry
  superalgebra of all these solutions.  We then construct various
  M-theory and string theory branes with homogeneous fluxes and we
  also find new homogeneous flux-brane solutions.
\end{abstract}

\maketitle

\section{Introduction}

Solutions of supergravity theory that preserve all spacetime
supersymmetry have played a central rôle in the understanding of the
properties of superstrings and M-theory.  The applications range from
the quantisation of superstrings in ten-dimensional Minkowski
spacetime to the formulation of the AdS/CFT correspondence.  Most of
our understanding is based on the large number of unbroken
supersymmetries that such solutions exhibit.  In M-theory, it is well
known that backgrounds that preserve all thirty-two supersymmetries
include eleven-dimensional Minkowski space, $\AdS_4\times S^7$ and
$\AdS_7\times S^4$.  Less well-known is that in addition to the above
three spaces there is another solution of $D=11$ supergravity theory
which preserves all supersymmetries.  This solution was discovered by
Kowalski-Glikman\cite{KG} and in what follows we shall refer to it as
the KG space or solution.

In this paper, we shall investigate solutions of supergravity theories
in ten and eleven dimensions that have covariantly constant form
field-strength fluxes.  In this paper we will call these fluxes
\emph{homogeneous}, since the spacetimes we will be concerned with
will be symmetric spaces, and for a symmetric space these two concepts
agree.  We therefore define a homogeneous $p$-brane (or
\emph{H$p$-brane}) to be a solution with $\ISO(p,1)$ Poincaré
invariance and with a U-duality representative which has homogeneous
fluxes.  Such solutions do not necessarily carry usual $p$-brane
charges but as we shall see in some cases describe the asymptotic
geometry of $p$-branes in the presence of a homogeneous flux.  We
shall also find that our H-branes are closely related to flux-branes
(or F-branes).  In fact, we shall show that if instead of reducing
Minkowski spacetime \cite{DGGH1, DGGH2}, we reduce a certain H-brane
spacetime, we obtain an F-brane in the presence of additional fluxes.

We shall focus on the H-branes that are constructed from a large
family of eleven-dimensional Lorentzian symmetric spacetimes found by
Cahen and Wallach \cite{CahenWallach}.  We call them Cahen--Wallach
(CW) spaces.\footnote{No confusion should arise with the unrelated
  notion of a CW complex, familiar to topologists.}  In particular we
shall show that generic CW spaces equipped with a homogeneous
four-form null flux are solutions of eleven-dimensional supergravity
preserving one half of the supersymmetry.  The resulting solution is
an M-theory pp-wave in homogeneous four-form flux, or Hpp-wave.  Our
solutions include a spacetime with homogeneous fluxes found in
\cite{RT}.

We shall then investigate the moduli space of these solutions and we
shall find that there are two special points.  One is
eleven-dimensional Minkowski spacetime and the other is the KG
spacetime.  At both of these points there is supersymmetry enhancement
and the solutions preserve all thirty-two supersymmetries of M-theory.

We shall find the Killing vectors and Killing spinors, and compute the
symmetry superalgebra of Hpp-waves.  The Killing vectors and Killing
spinors of the KG solution were computed explicitly in \cite{CKG}.
Anticipating the importance that the KG solution may have in the
context of M-theory, we shall compute its symmetry superalgebra.  We
shall find that the dimension of the bosonic part of the algebra of
the KG solution is $38$ which is the same as that of the $\AdS_4\times
S^7$ and $\AdS_7\times S^4$ solutions of M-theory.  However unlike the
latter two cases where the isometry group is semisimple, the isometry
group of KG solution is not.  The symmetry superalgebra of the rest of
the Hpp-waves will also be given.

We shall investigate the reduction of the M-theory Hpp-wave that we
have found and in particular that of the KG spacetime to IIA string
theory. The reduced solution is a IIA H0-brane. It is known that in
some cases after Kaluza--Klein reduction, the reduced solution
preserves less supersymmetry than the original one, see for example
\cite{Bakas, BKODuality, LPTAdS}.  It has been argued in \cite{GGPT}
for the special case of T-duality on toric hyperkähler manifold, but
applies more generally, that a necessary condition for preserving
supersymmetry in supergravity theory is that the Lie derivative of the
Killing spinors along the direction of compactification vanishes.
This condition does not depend on the choice of coordinates and the
choice of frame of spacetime.  More generally, the number of
supersymmetries preserved in supergravity theory after reducing a
supersymmetric solution is equal to the number of Killing spinors
which have vanishing Lie derivatives along the direction of
compactification.  We shall see that the above condition on the
Killing spinors is equivalent to requiring that the supercharges
associated with the reduced solution are those of the original
supersymmetric solution that commute with the generator of
translations along the compact direction.  In particular we find that
for the generic Hpp-wave the radius of the compact direction can be
chosen such that the solution preserves one half of the supersymmetry
in eleven dimensions.  However, the Lie derivative along the compact
direction of the Killing spinors of an Hpp-wave does not vanish.
Therefore, the associated H0-brane does not preserve any supersymmetry
in the context of IIA supergravity.  This is also reminiscent of the
backgrounds with supersymmetry without supersymmetry of \cite{DLP}.
For the Hpp-wave associated with the KG spacetime, the radius of the
compact direction can be chosen such that the solution preserves
thirty-two or sixteen supersymmetries in eleven dimensions.  However
again, the Lie derivative of the Killing spinors does not vanish along
the compact direction and so the associated H0-brane does not preserve
any supersymmetry in IIA supergravity.  Using U-duality one can
construct H$p$-branes in IIA ($p$ even) and IIB ($p$ odd)
supergravities.

We shall also consider the superposition of our H-brane solutions with
the standard M-branes.  We shall in fact focus on the superposition of
an M-theory pp-wave \cite{MWave} with an Hpp-wave.  We shall find that
upon reduction to IIA supergravity, the solution has the
interpretation of D0-branes in the presence homogeneous fluxes, or
HD0-branes.  Using U-duality, one can then construct HD$p$- and
HNS$p$-branes in type II theories.

Our H-brane solutions can also be used to construct a class of
flux-branes with additional homogeneous fluxes (\emph{HF-branes}, for
short).  For this, we shall perform a reduction in a class of our
M-theory Hpp-wave solutions.  The resulting IIA configuration has a
magnetic two-form flux characteristic of a F7-brane, in addition to
the fluxes which are reductions of the homogeneous four-form field
strength.

This paper has been organised as follows.  In Section~\ref{sec:CW}, we
summarise the construction of CW spaces and investigate their
geometry.  In Section~\ref{sec:HppCW} we give the Hpp-wave solutions
of eleven-dimensional supergravity.  In Section~\ref{sec:spinors} we
compute the Killing spinors for the Hpp-waves and those of KG
spacetime. In Section~\ref{sec:isometries} we examine the isometries
of the Hpp-waves. In Section~\ref{sec:superalgebra} we give the
symmetry superalgebra of Hpp-waves and in particular that of KG
spacetime.  In Section~\ref{sec:hbranes} we use U-duality to construct
many H-brane solutions.  In Section~\ref{sec:susiesH0} we use the
spinorial Lie derivative to investigate the number of supersymmetries
preserved by a solution after a Kaluza--Klein reduction. We then apply
our results in the context of H-branes.  In Section~\ref{sec:HDHNS} we
find HD- and HNS- brane solutions.  In Section~\ref{sec:HF} we find
HF-brane solutions. Finally, in Section~\ref{sec:conc}, we discuss the
moduli space of maximally supersymmetric solutions of
eleven-dimensional supergravity.

\section{Cahen--Wallach Lorentzian Symmetric Spaces}
\label{sec:CW}

In this section we present the construction due to Cahen and Wallach
\cite{CahenWallach} of a family of indecomposable Lorentzian symmetric
spaces.  This family exists for any dimension $\geq 3$, but we will
consider only the eleven-dimensional case---the general case following
straightforwardly from this one.  Let $x^\pm$ be light-cone
coordinates, ${x^i; i=1,\dots, 9}$ be coordinates in $\RR^9$ and
$A=(A_{ij})$ be a real symmetric matrix.  The CW metric is
\begin{equation}
  \label{eq:metric}
  ds^2 = 2 dx^+ dx^- + \sum_{i,j} A_{ij} x^i x^j
  (dx^-)^2 +  \sum_i dx^i dx^i
\end{equation}
It is not hard to show that the resulting space is indecomposable
(that is, it is not locally isometric to a product) if and only if $A$
is non-degenerate.  On the other hand if $A$ is degenerate then the
metric \eqref{eq:metric} decomposes to a $(11-k)$-dimensional
indecomposable CW space and the standard metric on $\RR^k$, where $k$
is the dimension of the radical of $A$.

In order to make the dependence of the CW metric on $A$ explicit, we
will let $M_A$ denote the space with CW metric corresponding to the
matrix $A$.  Some of the CW spaces associated with different matrices
$A$ are isometric.  To be precise, two CW spaces $M_{A_1}$ and
$M_{A_2}$ are (locally) isometric if and only if $A_1=c^2 O A_2
O^{-1}$, where $c$ is a non-vanishing (real) constant and $O$ is an
orthogonal transformation.  Indeed, the diffeomorphism is given
explicitly by rotating the $x^i$ by $O$ and then rescaling $x^\pm
\mapsto c^{\pm1} x^\pm$.

The moduli space of eleven-dimensional CW metrics is that of the
unordered eigenvalues of $A$ up to a positive scale.  This space is
diffeomorphic to $S^8/\Sigma_9$ where $\Sigma_9$ is the permutation
group of nine objects acting in the obvious way on the homogeneous
coordinates of $S^8$, that is, on the eigenvalues $(\lambda_1, \dots,
\lambda_9)$ of the matrix $A$.  The moduli space of indecomposable CW
metrics is
\begin{equation*}
  (S^8 \backslash \Delta)/\Sigma_9 \qquad
  \text{where $\Delta = \{(\lambda_1,\dots,\lambda_9)\in S^8 \mid
  \lambda_1 \lambda_2 \cdots \lambda_9 = 0\}$},
\end{equation*}
which is noncompact.  Adding the decomposable CW metrics compactifies
the moduli space to $S^8/\Sigma_9$.  This provides a simple example of
a phenomenon that has been observed in the compactification of many
other moduli spaces.

To show that the metric \eqref{eq:metric} is complete, to find the
isometries and to understand the global structure of $M_A$, we shall
construct $M_A$ as a symmetric space.  To this end, let $V$ be a real
$9$-dimensional vector space and let $V^*$ be the dual space.  We may
identify $V$ and $V^*$ with $\RR^9$ by choosing a basis $\{e_i\}$ for
$V$ and canonical dual basis $\{e_i^*\}$ for $V^*$.  We define a
euclidean inner product on $V$ by declaring the $e_i$ to be an
orthonormal basis.  Let $Z$ be a one-dimensional real vector space and
let $Z^*$ be its dual space.  We identify them with $\RR$ by choosing
canonically dual bases $e_+$ for $Z$ and $e_-$ for $Z^*$.  Finally let
$A$ be a real symmetric bilinear form on $V$: $A(e_i,e_j) = A_{ij} =
A_{ji}$.  Then the $20$-dimensional vector space $V\oplus V^* \oplus Z
\oplus Z^*$ can be made into a solvable Lie algebra with the following
nonzero Lie brackets:
\begin{equation}
  \label{eq:liealg}
  [e_-,e_i] = e^*_i \qquad [e_-,e^*_i] = \sum_j A_{ij} e_j \qquad
  [e^*_i,e_j] = A_{ij} e_+~.
\end{equation}
The brackets satisfy the Jacobi identity by virtue of $A$ being
symmetric.  We call the resulting Lie algebra $\fg_A$, since it
depends on $A$.  Notice that the second derived ideal $\fg''_A$ is
central, whence $\fg_A$ is solvable.  The Lie algebra
\eqref{eq:liealg} is isomorphic to a Heisenberg algebra generated by
$\{e_i, e_i^*, e_+\}$ and equipped with an outer automorphism $e_-$
which rotates positions $\{e_i\}$ and momenta $\{e_i^*\}$.  Systems
with such symmetry include the harmonic oscillator, the Planck
``constants'' being given by the eigenvalues of $A$.

Let $\fk_A$ denote the abelian Lie subalgebra spanned by the
$\{e_i^*\}$ and let $\fp_A$ denote the complementary subspace, spanned
by $\{e_i,e_+,e_-\}$.  It is clear from the Lie brackets
\eqref{eq:liealg} that
\begin{equation*}
  [\fk_A,\fp_A] \subset \fp_A \qquad\text{and}\qquad
  [\fp_A,\fp_A] \subset \fk_A~,
\end{equation*}
whence $\fg_A = \fk_A \oplus \fp_A$ is a symmetric split.
Furthermore, the action of $\fk_A$ on $\fp_A$ preserves the symmetric
bilinear form
\begin{equation}
  \label{eq:sbf}
  B(e_i,e_j) = \delta_{ij} \qquad\text{and}\qquad
  B(e_+,e_-) = 1~.
\end{equation}

Let $G_A$ denote the simply-connected Lie group with Lie algebra
$\fg_A$ and let $K_A$ denote the Lie subgroup corresponding to the Lie
subalgebra $\fk_A$.  Then the space $M_A = G_A/K_A$ of right
$K_A$-cosets in $G_A$ inherits a Lorentzian metric from the
$K_A$-invariant bilinear form $B$ on $\fp_A$.  In this way, $M_A$
becomes a Lorentzian symmetric space with solvable transvection group
$G_A$.

To express the metric on $M_A$ as in \eqref{eq:metric}, we choose a
representative $\sigma: M_A \to G_A$ of the coset as
\begin{equation}
  \label{eq:cosetrep}
  \sigma(x^+, x^-, x^i) = \exp(x^+ e_+) \exp(x^- e_-) \exp\left(\sum_i
    x^i e_i\right)~,
\end{equation}
where $\{x^+,x^-,x^i\}$ are local coordinates as in \eqref{eq:metric}.
(Normally this can only be done locally, but in this case, since $M_A$
is contractible, we can choose a global representative.)  The
pull-back to $M_A$ of the left-invariant Maurer--Cartan form on $G_A$
can be written as
\begin{equation*}
  \sigma^{-1} d\sigma = \omega + \theta~,
\end{equation*}
where the canonical $\fk_A$-connection $\omega$ of the coset and the
$\fp_A$-valued soldering form $\theta$ (whose components give rise to
a frame) are given by
\begin{equation*}
  \omega = dx^- \sum_i x^i e^*_i
\end{equation*}
and
\begin{equation*}
  \theta = dx^- e_- + \sum_i dx^i e_i + \left(dx^+ + \half \sum_{i,j}
  A_{ij}x^i x^j dx^-\right) e_+~,
\end{equation*}
respectively.  In the above coordinates, the $G_A$-invariant metric
$B(\theta, \theta)$ on $M_A$ coincides with \eqref{eq:metric}.

For later use, the Riemann curvature and Ricci tensors of $M_A$ have
the following nonzero components
\begin{equation}
  \label{eq:curvature}
  R_{-i-j}=-A_{ij} \qquad\text{and}\qquad
  R_{--}= -\tr A~.
\end{equation}
In particular, the scalar curvature vanishes.

The holonomy of $M_A$ is isomorphic to $K_A \cong \RR^9$ and the
representation is induced by the adjoint action of $K_A$ on $\fp_A$.
It is not hard to show that the $K_A$-invariant subspace of $\bigwedge
\fp_A$ is spanned by the constants together with polyvectors of the
form $e_+ \wedge \theta$, where $\theta \in \bigwedge V$.  This means
that the corresponding parallel forms on $M_A$ are the constants
together with $dx^- \wedge \Theta$ where $\Theta$ is any
constant-coefficient form
\begin{equation*}
\label{eq:formm}
  \Theta = \sum_{1\leq i_1<i_2<\cdots<i_p \leq 9} c_{i_1i_2\cdots
  i_p} dx^{i_1} \wedge dx^{i_2} \wedge \cdots \wedge dx^{i_p}~.
\end{equation*}

\section{Hpp-waves from CW Spaces}
\label{sec:HppCW}

The CW spaces $M_A$ described in the previous section can be used to
construct Hpp-wave solutions of eleven-dimensional supergravity.  The
bosonic fields of eleven-dimensional supergravity are the metric $g$
and the four-form field-strength $F$.  Since we want the Hpp-wave to
have the symmetries of the CW spaces, it is natural to demand that $F$
be homogeneous, so that we shall choose $F$ to be a parallel form on
the CW space.  In particular, we shall show that
\begin{equation}
  \label{eq:cwsoll}
  \begin{aligned}[m]
    ds^2 &= 2 dx^+ dx^- + \sum_{i,j} A_{ij} x^i x^j (dx^-)^2 +  \sum_i
    dx^i dx^i\\
    F &=  dx^- \wedge \Theta,
  \end{aligned}
\end{equation}
where $\Theta$ is a 3-form on $\RR^9$ with constant coefficients, is a
supersymmetric solution of eleven-dimensional supergravity provided
that
\begin{equation}
  \label{eq:conn}
  \tr A = -\half \|\Theta\|^2=- \tfrac1{12} \Theta_{ijk}
  \Theta^{ijk}~.
\end{equation}

In the conventions of \cite{DS2brane}, the field equations of
eleven-dimensional supergravity  \cite{CJS} are
\begin{equation}
  \label{eq:eqns}
  \begin{aligned}[m]
    d\star F &= \half F \wedge F\\
    R_{MN} &= \tfrac1{12} \left( F_{MPQR} F_N{}^{PQR} - \tfrac1{12}
      g_{MN} F_{PQRS} F^{PQRS}\right)~.
  \end{aligned}
\end{equation}

For later use, we also give the Killing spinor equations of
eleven-dimensional supergravity.  Let $\{\Gamma_a\}$ be a basis in the
Clifford algebra
\begin{equation*}
  \Gamma_a \Gamma_b + \Gamma_b \Gamma_a =  2 \eta_{ab} \1~,
\end{equation*}
where $\eta_{ab}$ is the ``mostly plus'' frame metric.  A convenient
basis for investigating Killing spinors for the solution
\eqref{eq:cwsol} is one for which $\eta_{+-} = 1$ and $\eta_{ij} =
\delta_{ij}$.  The Killing spinors $\varepsilon$ satisfy the equation
\begin{equation}
  \label{eq:Killing}
  \nabla_M \varepsilon = \Omega_M \varepsilon~,
\end{equation}
where
\begin{equation}
  \label{eq:Omega}
  \Omega_M = \tfrac1{288}  F_{PQRS}
  \left( \Gamma^{PQRS}{}_M + 8 \Gamma^{PQR}\delta^S_M \right)~,
\end{equation}
and the spin connection $\nabla$ is 
\begin{equation}
  \label{eq:spinconn}
  \nabla_M = \d_M + \tfrac14 \omega_M{}^{ab} \Gamma_{ab}~.
\end{equation}

To see that \eqref{eq:cwsoll} satisfies the field equations of
eleven-dimensional supergravity theory, we remark that $F$ is
parallel, hence it is both closed and coclosed.  Moreover it obeys
$F\wedge F = 0$.  Therefore $F$ obeys its equation of motion.  Since
$F$ is null---that is, $F_{PQRS}F^{PQRS} = 0$---the Einstein equations
simplify to
\begin{equation*}
  R_{MN} = \tfrac1{12} F_{MPQR} F_N{}^{PQR}~.
\end{equation*}
The only nonzero component of the Ricci tensor is $R_{--} = -\tr A$,
and similarly the only nonzero component of energy-momentum tensor
$T_{MN} := F_{MPQR} F_N{}^{PQR}$ is $T_{--} = 6 \|\Theta\|^2$, whence
we see that in order to obtain a bosonic solution, all we need to
impose is the condition \eqref{eq:conn}.

It is clear from our construction that the solution \eqref{eq:cwsoll}
is invariant under the action of the Lie algebra $\fg_A$ of the
transvection group and that it has the interpretation of an Hpp-wave
according to the definition given in the introduction.  The Hpp-wave
\eqref{eq:cwsoll} may be invariant under the action of a larger group.
A detailed investigation of the symmetries will be presented in
Sections~\ref{sec:isometries} and \ref{sec:superalgebra}.  Here we
shall simply mention that the Hpp-wave \eqref{eq:cwsoll} is invariant
under the action of the Lie algebra $\fg_A \rtimes (\fs_A \cap
\fs_\Theta)$, where $\fs_A \subset \fso(V)$ and $\fs_\Theta \subset
\fso(V)$ are the isotropy algebras of $A$ and $\Theta$, respectively,
and $\rtimes$ is the semidirect sum.   The action of $\fs_A \cap
\fs_\Theta$ on $\fg_A$ is induced by restriction from the natural
action of $\fso(V)$ on $V\oplus V^* \subset \fg_A$.  In the next
section we shall show that generic Hpp-wave solutions preserve one
half of the supersymmetry.

There are some special cases of Hpp-waves that we can consider.  One
possibility is to choose $\Theta=0$.  In such case, the spacetime is
not necessarily Minkowski.  The condition \eqref{eq:conn} only implies
that the trace of $A$ vanishes.  The associated CW spaces are
\emph{Ricci flat} but not isometric to Minkowski space.  However if
$A=0$, we recover eleven-dimensional Minkowski space.  Therefore the
moduli space of Hpp-waves has a point that preserves thirty-two
supersymmetries.  As we will see presently, it has exactly one other
such point.

Another case that we shall  focus later is to take
\begin{equation}
  \label{eq:cwsol}
  \begin{aligned}[m]
    ds^2 &= 2 dx^+ dx^- + \sum_{i,j} A_{ij} x^i x^j
    (dx^-)^2 +  \sum_i dx^i dx^i\\
    F&= \mu\, dx^- \wedge dx^1 \wedge dx^2 \wedge dx^3~,
  \end{aligned}
\end{equation}
where $\mu$ is a constant.  In such a case the equation \eqref{eq:conn}
implies that
\begin{equation}
  \label{eq:con}
  \tr A = -\half \mu^2~.
\end{equation}
The isotropy algebra of $F$ is $\fso(3)\oplus\fso(6)$.  We shall show
in Section~\ref{sec:spinors} that for a special choice of $A$, this
solution preserves thirty-two supersymmetries.  This is the KG
solution of eleven-dimensional supergravity.

Another special case of \eqref{eq:cwsol} is the solution given in
\cite{RT}.  This solution corresponds to a reducible CW space.  To see
this, we write the metric given in \cite{RT} as
\begin{equation*}
  ds^2= -dt^2 + dx_9^2 + 2 b r^2 d\varphi (dx_9 - dt) + dr^2 + r^2
  d\varphi^2 + ds^2(\RR^7)~.
\end{equation*}
Next we change coordinates
\begin{equation*}
  x^\pm=\frac1{\sqrt{2}} (\pm t+x_9) \qquad \tilde \varphi=
  \varphi+b(x_9-t)~.
\end{equation*}
In these new coordinates the metric is written as
\begin{equation*}
 ds^2 = 2 dx^- dx^+ - 2 b^2 r^2 (dx^-)^2 + dr^2 + r^2 d\tilde
 \varphi^2 + ds^2(\RR^7)~.
\end{equation*} 
After periodically identifying $\tilde \varphi$ and changing from
polar $(r,\tilde\varphi)$ to rectangular coordinates on $\RR^2$, the
metric we find is that of a decomposable CW space for which $A$ has
two equal non-vanishing eigenvalues $\lambda_1=\lambda_2=-2b^2$.

\section{Killing Spinors}
\label{sec:spinors}

There are two cases to consider.  First we shall show that generic
Hpp-waves \eqref{eq:cwsoll} preserve one half of the supersymmetry,
that is, that they admit sixteen linearly independent Killing spinors.
Then we shall prove that there is a special point in the moduli space
of these solutions, which is not Minkowski space, where supersymmetry
is enhanced and the solution admits thirty-two linearly independent
Killing spinors.  In particular, we shall show that if
\begin{equation}
  \label{eq:Aij}
  \begin{aligned}[m]
    A_{ij}& = 
    \begin{cases}
      -\tfrac19 \mu^2 \delta_{ij} & i,j=1,2,3\\
      -\tfrac1{36} \mu^2 \delta_{ij} & i,j=4,5,\dots,9
    \end{cases}\\
    \Theta&=\mu dx^1\wedge dx^2\wedge dx^3
  \end{aligned}
\end{equation}
then the solution \eqref{eq:cwsol} preserves all supersymmetry.  This
is the KG vacuum solution of eleven-dimensional supergravity theory.

To begin the analysis, the nonvanishing components of the spin
connection one-form of the \eqref{eq:cwsol} solution are
\begin{equation*}
  \omega^{+i} = - \omega^{i+} = \sum_j A_{ij} x^j dx^-~,
\end{equation*}
where $+i$ are frame indices.  In addition, using that $F_{-123}=\mu$
and all other components vanish, the $F$-dependent piece $\Omega_M$ of
the Killing spinor equation \eqref{eq:Killing} has the following
nonzero components\footnote{Throughout this section, the indices on
  the $\Gamma$-matrices are frame indices.}
\begin{equation}
  \label{eq:Omegaexplicit}
  \begin{aligned}[m]
    \Omega_- &= - \tfrac1{12} \mu \left( \Gamma_+ \Gamma_- + \1
    \right) I\\
    \Omega_i &= 
    \begin{cases}
      \phantom{-}\tfrac16 \mu \Gamma_+ \Gamma_i I & i=1,2,3\\
      - \tfrac1{12} \mu \Gamma_+ \Gamma_i I  & i=4,5,\dots,9~,
    \end{cases}
  \end{aligned}
\end{equation}
where $I= \Gamma_{123}$ obeys $I^2 = - \1$.  It follows from
$\Gamma_+^2=0$ that
\begin{equation}
  \label{eq:omegaomega}
  \Omega_i \Omega_j = 0 \qquad\text{for all $i,j = 1,2,\dots,9$.}
\end{equation}

To solve the Killing spinor equations for the (generic) solution
\eqref{eq:cwsoll}, we impose the condition
\begin{equation}
  \label{eq:kilpro}
  \Gamma_+\varepsilon=0~.
\end{equation}
In this case, the Killing spinor equations reduce to
\begin{equation*}
  \d_- \varepsilon = \tfrac1{24} \Theta_{ijk} \Gamma^{ijk}
  \varepsilon~,
\end{equation*}
where $\varepsilon=\varepsilon(x^-)$ is only a function of $x^-$.
This equation can be solved and the Killing spinors are
\begin{equation}
\label{eq:hppspinora}
  \varepsilon = \exp\left(\tfrac1{24} x^- \Theta_{ijk}
  \Gamma^{ijk}\right)\psi_+~,
\end{equation}
for some constant spinor $\psi_+$ such that $\Gamma_+\psi_+=0$.  It is
clear from this that the generic solution \eqref{eq:cwsoll} preserves
one half of the supersymmetry.

In particular for the solution \eqref{eq:cwsol}, the Killing spinor
equations are
\begin{equation}
\label{eq:hppspinorb}
  \partial_-\varepsilon + \tfrac{\mu}{4} I\varepsilon=0~,
\end{equation}
and the Killing spinors are given as
\begin{equation*}
  \varepsilon= \left(\cos(\tfrac{\mu}{4} x^-) \1-\sin(\tfrac{\mu}{4}x^-)
  I\right) \psi_+~,
\end{equation*}
where $\psi_+$ is again a constant spinor satisfying
$\Gamma_+\psi_+=0$.

Next we turn to find the special point in the moduli space of
\eqref{eq:cwsol} solutions that exhibits enhancement of supersymmetry.
Since $\nabla_+ = \d_+$ and $\Omega_+ = 0$, we see that the Killing
spinor $\varepsilon$ is independent of $x^+$.  Similarly from
\begin{equation}
  \label{eq:killi}
  \d_i \varepsilon = \Omega_i \varepsilon
\end{equation}
and equation \eqref{eq:omegaomega}, we see that $\d_i \d_j \varepsilon
= 0$, whence $\varepsilon$ is at most linear in $x^i$.  Let us write
it as
\begin{equation*}
  \varepsilon = \chi + \sum_i x^i \varepsilon_i~,
\end{equation*}
where the spinors $\chi$ and $\varepsilon_i$ are only functions of
$x^-$.  From equation \eqref{eq:killi} we see that $\varepsilon_i =
\Omega_i \chi$, whence any Killing spinor $\varepsilon$ takes the form
\begin{equation}
  \label{eq:kill0}
  \varepsilon = \left(\1 + \sum_i x^i \Omega_i \right) \chi~,
\end{equation}
where the spinor $\chi$ only depends on $x^-$.  The dependence on
$x^-$ is fixed from the one remaining equation
\begin{equation*}
  \d_- \varepsilon = - \half \sum_{i,j} A_{ij} x^j \Gamma_+ \Gamma_i
  \varepsilon - \tfrac1{12} \mu \left( \Gamma_+ \Gamma_- + \1 \right)
  I \varepsilon~,
\end{equation*}
which will also imply an integrability condition fixing $A$.

Inserting the above expression \eqref{eq:kill0} for $\varepsilon$ into 
this equation and after a little bit of algebra (using repeatedly that
$\Gamma_+^2 = 0$), we find
\begin{multline}
  \label{eq:xks}
  \frac{d}{dx^-}\chi = - \tfrac1{12} \mu I \left(\1 + \Gamma_+
    \Gamma_-\right) \, \chi\\
  + \sum_i x^i \left( - \half \sum_j A_{ij} \Gamma_+ \Gamma_j +
    \tfrac1{12} \mu \Omega_i I - \tfrac14 \mu I \Omega_i\right)\,
  \chi~.
\end{multline}

Because $\chi$ is independent of $x^i$, the terms in parenthesis in
the right-hand side of the equation must vanish separately.  This will
fix $A$.  The remaining equation is a first-order linear ordinary
differential equation with constant coefficients, which has a unique
solution for each initial value.  Since the initial value is an
arbitrary spinor, we see that the dimension of the space of Killing
spinors is $32$ and hence the solution will be maximally
supersymmetric.

To fix $A$, notice that
\begin{equation*}
  \Omega_i I = 
  \begin{cases}
    -\tfrac16 \mu \Gamma_+ \Gamma_i & i=1,2,3\\
    \tfrac1{12} \mu \Gamma_+ \Gamma_i & i=4,5,\dots,9
  \end{cases}
\end{equation*}
and
\begin{equation*}
  I \Omega_i =
  \begin{cases}
    \tfrac16 \mu \Gamma_+ \Gamma_i & i=1,2,3\\
    \tfrac1{12} \mu \Gamma_+ \Gamma_i & i=4,5,\dots,9
  \end{cases}
\end{equation*}
whence the $x^i$-dependent terms in \eqref{eq:xks} vanish provided
that
\begin{equation*}
  \sum_j A_{ij} \Gamma_j = 
  \begin{cases}
    - \tfrac19 \mu^2 \Gamma_i & i=1,2,3\\
    - \tfrac1{36} \mu^2 \Gamma_i & i=4,5,\dots,9~,
  \end{cases}
\end{equation*}
whence $A$ is diagonal with eigenvalues $\lambda_i = - \frac19 \mu^2$
for $i=1,2,3$ and $\lambda_i = -\frac1{36} \mu^2$ for $i=4,5,\dots,9$.
In particular it is nondegenerate, whence the metric \eqref{eq:metric}
is indecomposable.  As a final check, notice that $\tr A = - \half
\mu^2$ as required from the equations of motion.  We conclude
therefore that the solution \eqref{eq:cwsol} with $A$ given in
\eqref{eq:Aij} is a maximally supersymmetric solution of
eleven-dimensional supergravity.  Notice that all nonzero values of
$\mu$ are isometric (simply rescale $x^-$), whereas $\mu=0$
corresponds to eleven-dimensional Minkowski spacetime.

One can give a more explicit expression for the Killing spinors.
Decompose the spinor $\chi$ as $\chi_+ + \chi_-$, where
$\Gamma_\pm\chi_\pm=0$.  Since $I$ preserves $\ker\Gamma_\pm$, the
first-order equation for $\chi$ breaks up into two equations:
\begin{equation*}
  \frac{d}{dx^-}\chi_+= - \tfrac14 \mu I \chi_+
  \qquad\text{and}\qquad
  \frac{d}{dx^-}\chi_- = - \tfrac1{12} \mu I \chi_-~,
\end{equation*}
which can be solved via matrix exponentials in terms of constant
spinors $\psi_\pm \in \ker \Gamma_\pm$:
\begin{equation*}
  \begin{aligned}[m]
    \chi_+ &= \exp\left(- \tfrac14 \mu x^- I \right) \psi_+ =
    \left( \cos(\tfrac14 \mu x^-)\1 - \sin(\tfrac14 \mu x^-) I\right)
    \psi_+\\
    \chi_- &= \exp\left(- \tfrac1{12} \mu x^- I \right) \psi_- =
    \left( \cos(\tfrac1{12} \mu x^-)\1 - \sin(\tfrac1{12} \mu x^-) I\right)
    \psi_-~.
  \end{aligned}
\end{equation*}
Finally, we use equation \eqref{eq:kill0} to arrive at the following
expression for $\varepsilon$ in terms of the arbitrary constant
spinors $\psi_\pm$:
\begin{footnotesize}
  \begin{multline}
    \label{eq:killingspinors}
    \varepsilon = \left( \cos(\tfrac14 \mu x^-)\1 - \sin(\tfrac14 \mu
      x^-) I\right) \psi_+ + \left( \cos(\tfrac1{12} \mu x^-)\1 -
      \sin(\tfrac1{12} \mu x^-) I\right) \psi_-\\
    - \tfrac16 \mu \left( \sum_{i\leq3} x^i \Gamma_i - \half \sum_{i\geq
        4} x^i \Gamma_i \right) \left( \sin(\tfrac1{12} \mu x^-)\1 -
      \cos(\tfrac1{12} \mu x^-) I\right) \Gamma_+ \psi_-~.
  \end{multline}
\end{footnotesize}
Observe that the Killing spinors depend trigonometrically on $x^-$.
These expressions agree (up to notation) with those found in
\cite{CKG} for the KG solution.

\section{Isometries}
\label{sec:isometries}

We shall begin by investigating the isometries of a generic CW space
$M_A$ and then we shall specialise to those of the KG solution.  Since
$M_A = G_A/K_A$ is the space of left cosets of $K_A$ in $G_A$, it
admits a natural action of $G_A$.  By construction, the metric on
$M_A$ is invariant under this action and hence the associated
transformations are isometries.  This means that there are $20$
linearly independent Killing vector fields $\xi_X$ each associated
with an element $X$ of the Lie algebra $\fg_A$ of $G_A$.  The map $X
\mapsto \xi_X$ is an isomorphism of Lie algebras, where the Lie
bracket of $\xi_X$ is the Lie bracket of vector fields on $M_A$.  To
express these Killing vector fields in terms of the coordinates
$\{x^\pm, x^i\}$, we shall use the local coset representative $\sigma$
given by \eqref{eq:cosetrep} and act on it with $G_A$ from the left.
The resulting right-invariant vector fields project to Killing vectors
on $M_A$ relative to the $G_A$-invariant metric.  To simplify the
expressions for the Killing vector fields, we will assume that the
coordinates $\{x^i\}$ have been chosen in such a way that $A$ is
diagonalised: $A_{ij} = \lambda_i \delta_{ij}$.

Let $X \in \fg_A$ and let $t\mapsto \exp t X$ be a one-parameter
subgroup of $G_A$ with tangent vector $X$ at the identity.  Acting
with this one-parameter subgroup on $\sigma(x)$, we obtain
\begin{equation*}
  \exp(t X) \, \sigma(x) = \sigma\left( \exp(t X) \cdot x\right)
  k(t,x)~,
\end{equation*}
for some $k(t,x) \in K_A$.  Notice that $k(0,x)$ is the identity for
all $x$, hence differentiating $k(t,x)$ with respect to $t$ at
$t=0$ we obtain an element ($Y$, say) of $\fk_A$.  Differentiating the 
previous equation with respect to $t$ at $t=0$ we therefore obtain
(abusing notation slightly)
\begin{equation*}
  X \sigma(x) =\xi_X + \sigma(x) Y~,
\end{equation*}
where $Y\in\fk_A$ and where $\xi_X$ is the local coordinate
expression for the Killing vector corresponding to $X$.  In other
words,
\begin{equation*}
  \xi_X = X \sigma(x) \pmod {\sigma(x) \fk_A}~.
\end{equation*}
To express $\xi_X$ in terms of the coordinate vectors
$\{\d_\pm,\d_i\}$, we need to recognise these in $G_A$.  Recalling
that the action on coordinates is inverse to that on points, the
vector fields $\d_+$, $\d_-$ and $\d_i$ at the point with coordinates
$\{x^+, x^-, x^j\}$ are the tangent vectors, respectively, to the
following curves on $M_A$.
\begin{equation*}
  t \mapsto (x^+ - t, x^-, x^j) \quad t \mapsto (x^+, x^- - t, x^j)
  \quad t \mapsto (x^+, x^-, x^j - t \delta_{ij})~.
\end{equation*}
We can think of these as curves on $G_A$ by composing them with
$\sigma$.  Then their tangent vectors at the point $\sigma(x)$ become
(with a slight abuse of notation):
\begin{equation*}
  - e_+ \sigma(x) \qquad - e_- \sigma(x) \qquad - \sigma(x) e_i~,
\end{equation*}
which are the expressions in $G_A$ for (the push-forwards via $\sigma$
of) $\d_+$, $\d_-$ and $\d_i$, respectively.  Using the above
computations, we find that the Killing vector fields associated with
the action of $G_A$ on the coset $M_A$ are the following:
\begin{equation*}
  \begin{aligned}[m]
   \xi_{e_+}&=-\d_+ \qquad\qquad \xi_{e_-}= -\d_-\\
    \xi_{e_i} &= - C_i (x^-) \d_i + S_i (x^-) x^i \lambda_i \d_+\\
    \xi_{e^*_i} &= \lambda_i S_i (x^-) \d_i - C_i(x^-) x^i \lambda_i
    \d_+~,
  \end{aligned}
\end{equation*}
(there is no sum over $i$), where the functions $C_i (x^-)$ and
$S_i (x^-)$ are given by
\begin{equation*}
  C_i (x^-) = 
  \begin{cases}
    1 & \text{if $\lambda_i = 0$}\\
    \cosh\left( \sqrt{\lambda_i}\, x^- \right) & \text{if $\lambda_i > 0$}\\
    \cos\left( \sqrt{-\lambda_i}\, x^- \right) & \text{if $\lambda_i < 0$,}
  \end{cases}
\end{equation*}
and
\begin{equation*}
  S_i (x^-) = 
  \begin{cases}
    x^- & \text{if $\lambda_i = 0$}\\
    \frac{\sinh\left( \sqrt{\lambda_i}\, x^-
    \right)}{\sqrt{\lambda_i}} & \text{if $\lambda_i > 0$}\\
    \frac{\sin\left( \sqrt{-\lambda_i}\, x^-
    \right)}{\sqrt{-\lambda_i}} & \text{if $\lambda_i < 0$,}
  \end{cases}
\end{equation*}
respectively.  One can check that the Killing vector fields obtained
above do indeed form a representation of $\fg_A$.  We note that the
supergravity solution of Section~\ref{sec:HppCW} associated with the
CW space $M_A$ is also invariant under the above isometries; that is,
the above isometries leave invariant the four-form field-strength $F$
as well.

For generic $A$ these are all the Killing vectors of the metric
\eqref{eq:metric}, but if $A$ is invariant under some subalgebra
$\fs_A \subset \fso(V)$, then there are extra Killing vectors
corresponding to $\fs_A$.  Typically this occurs whenever two or more
eigenvalues of $A$ coincide.  The subalgebra $\fs_A$ is generated by
the rotations that leave invariant the eigenspace corresponding to
these eigenvalues.  The Lie algebra $\fs_A$ acts naturally on $\fg_A$
by restricting the action of $\fso(V)$ on $V \oplus V^* \subset
\fg_A$.  Since $\fs_A$ leaves $A$ invariant, it preserves the Lie
bracket on $\fg_A$ and hence we can define the semi-direct product
$\fg_A \rtimes \fs_A$.  It can be shown that this is the isometry
algebra of $M_A$.  The unenlightening proof (which we omit) involves
solving Killing's equation explicitly.

The symmetries of the Hpp-waves solution \eqref{eq:cwsoll} leave both
the eleven-dimensional metric and four-form field strength invariant.
Adapting the explanation we have presented above for $M_A$ it is easy
to see that the symmetry group of Hpp-waves is $\fg_A \rtimes
(\fs_A\cap \fs_\Theta)$, where $\fs_\Theta$ is the subalgebra of
$\fso(V)$ that leaves the three-form $\Theta$ invariant.  For example,
if $A_{ij}=\delta_{ij}$, then the algebra of isometries of the CW
space $M_I$ is $\fg_A \rtimes \fso(9)$.   For the associated Hpp-wave
\eqref{eq:cwsol}, however, the symmetry subalgebra is $\fg_A \rtimes
(\fso(3)\oplus \fso(6))$.

Now we shall investigate the symmetries of the maximally
supersymmetric KG solution, which was originally computed in
\cite{CKG}.  The eigenvalues of $A$ given by \eqref{eq:Aij} are
negative, hence just like the Killing spinors, the Killing vectors
coming from the $G_A$-action depend trigonometrically on $x^-$.  Apart
from the symmetries associated with $\fg_A$, this solution also has an
additional $\fs_A \cong \fso(3)\oplus \fso(6)$ invariance because $A$
has two distinct eigenvalues with three- and six-dimensional
eigenspaces, respectively.  The Killing vectors of the KG background
are the following:
\begin{equation}
  \label{eq:killingvectors}
  \begin{aligned}[m]
   \xi_{e_+}&=-\d_+ \qquad\qquad \xi_{e_-}= -\d_-\\
   \xi_{e_i} &=- \cos\left(\tfrac{\mu x^-}{3}\right) \d_i -
   \sin\left(\tfrac{\mu x^-}{3}\right) \tfrac{\mu x^i}{3} \d_+
   \qquad\text{$i=1,2,3$}\\
   \xi_{e^*_i} &=- \sin\left(\tfrac{\mu x^-}{3}\right) \tfrac{\mu}{3}
   \d_i + \cos\left(\tfrac{\mu x^-}{3}\right) \tfrac{\mu^2 x^i}{9} \d_+
    \qquad\text{$i=1,2,3$}\\
    \xi_{e_i}&=- \cos\left(\tfrac{\mu x^-}{6}\right) \d_i -
    \sin\left(\tfrac{\mu x^-}{6}\right) \tfrac{\mu x^i}{6} \d_+
        \qquad\text{$i=4,5,6,7,8,9$}\\
    \xi_{e^*_i}&=- \sin\left(\tfrac{\mu x^-}{6}\right) \tfrac{\mu}{6}
    \d_i + \cos\left(\tfrac{\mu x^-}{6}\right) \tfrac{\mu^2 x^i}{36}
    \d_+ \qquad\text{$i=4,5,6,7,8,9$}\\
    \xi_{M_{ij}}&= x^i \d_j-x^j \d_i \qquad\text{$i,j=1,2,3$}\\
    \xi_{M_{ij}}&=x^i \d_j-x^j \d_i \qquad\text{$i,j=4,5,6,7,8,9$,}
  \end{aligned}
\end{equation}
where $\{M_{ij};i,j=1,2,3\}$ and $\{M_{ij}; i,j=4,5,6,7,8,9\}$ are
generators of $\fso(3)$ and $\fso(6)$, respectively.

The bosonic generators of the symmetry algebra of the 
KG solution is $\fg_A \rtimes (\fso(3)
\oplus \fso(6))$ has dimension $38$, which (intriguingly) is of the
same dimension as the isometry algebras of the other nontrivial
maximally supersymmetric solutions: $\AdS_4 \times S^7$ and $\AdS_7
\times S^4$. 

\section{The symmetry superalgebra}
\label{sec:superalgebra}

In addition to the bosonic symmetries generated by Killing vector
fields, solutions of supergravity theories preserving some
supersymmetry are also invariant under fermionic symmetries generated
by shifts along their Killing spinors.  Combining the bosonic and
fermionic symmetries, such backgrounds are invariant under the action
of a supergroup.  In the case of backgrounds of the form $\AdS_p
\times X$, the precise form of the associated Lie superalgebra was
elucidated in \cite{JMFKilling,PKT}.  In this section we do the same
for the M-theory Hpp-waves of \eqref{eq:cwsoll}.  We shall first find
the symmetry superalgebra of the KG solution and then we shall give
the symmetry superalgebra of a generic Hpp-wave.

The Lie superalgebra in question will be denoted $\fS$, and it splits
into even and odd subspaces $\fS = \fS_0 \oplus \fS_1$.  The odd
subspace is spanned by the Killing spinors and the even subspace by
those Killing vectors also preserving $F$.  In order to define the
structure of a Lie superalgebra we need to construct linear maps
\begin{equation*}
  \begin{aligned}[m]
    [\cdot, \cdot] &: \fS_0 \times \fS_0 \to \fS_0\\
    [\cdot, \cdot] &: \fS_0 \times \fS_1 \to \fS_1\\
    \{\cdot, \cdot\} &: \fS_1 \times \fS_1 \to \fS_0~,
  \end{aligned}
\end{equation*}
subject to the Jacobi identities.  The first map is simply the Lie
bracket of vector fields and it clearly satisfies the Jacobi identity.
The second map is the action of Killing vectors on Killing spinors,
making the space of Killing spinors into a linear representation of
the isometry subalgebra.  As in \cite{JMFKilling,PKT}, this is
achieved by the spinorial Lie derivative.  If $X$ is a Killing vector,
then one can define a Lie derivative $L_X$ on a spinor $\psi$ as
\begin{equation*}
  L_X \psi= X^M \nabla_M \psi+ \tfrac14 \nabla_{[M} X_{N]}
  \Gamma^{MN}\psi~.
\end{equation*}
As it has been explained in \cite{JMFKilling}, this has the
following  properties:
\begin{enumerate}
\item If $f$ is any smooth function and $\psi$ is any spinor, then
  \begin{equation*}
    L_X (f\psi) = (Xf)\psi + f L_X \psi~;
  \end{equation*}
\item If $X$ is a Killing vector field, $Y$ is \emph{any} vector
  field, and $\psi$ any spinor, then
  \begin{equation*}
    L_X (Y \cdot \psi) = [X,Y] \cdot \psi + Y \cdot L_X \psi~,
  \end{equation*}
  where $\cdot$ denotes the Clifford action of vectors on spinors; and
\item If $X,Y$ are Killing vector fields and $\psi$ is a spinor,
  \begin{equation*}
    L_X L_Y \psi  - L_Y L_X \psi = L_{[X,Y]} \psi~.
  \end{equation*}
\end{enumerate}
Notice that the spinorial Lie derivative preserves the Spin-invariant
inner product on the space of spinors.

If $X$ is a Killing vector which in addition preserves the four-form
$F$, then it is easy to show that $L_X$ preserves the space of Killing
spinors $\fS_1$ of a eleven-dimensional background.  
Therefore $\fS_1$ becomes a representation of the
Lie algebra $\fS_0$: the Lie algebra spanned by Killing vectors which
preserve $F$.

Finally, the last map $\{\cdot, \cdot\}$ in the structure of the Lie
superalgebra $\fS$ is simply the squaring of Killing spinors.  Indeed,
it is easy to show that if $\varepsilon_i$, $i=1,2$ are Killing
spinors, then the vector field with components $\bar\varepsilon_1
\Gamma^M \varepsilon_2$ is a Killing vector.  For the background in
question, these Killing vectors also preserve $F$.  The second
property of the Lie derivative guarantees that this operation is
equivariant under the action of isometries, which is one component of
the Jacobi identity for the Lie superalgebra $\fS$.

We will now explicitly exhibit the symmetry superalgebra.  The
commutators of the bosonic part of the superalgebra are
\begin{footnotesize}
  \begin{equation}
    \label{eq:S0S0}
    \begin{aligned}[m]
      [e_-,e_i] &= e^*_i \qquad [e_-,e^*_i] = - \tfrac{\mu^2}{9}
      e_i\quad (i\leq3) \qquad [e_-,e^*_i] = - \tfrac{\mu^2}{36}
      e_i\quad (i\geq4)
      \\
      [e^*_i,e_j]& = -\tfrac{\mu^2}{9} e_+\quad (i,j\leq3) \qquad
      [e^*_i,e_j]
      = -\tfrac{\mu^2}{36} e_+ \quad (i,j\geq4)\\
      [M_{ij}, e_k]&= -\delta_{ik} e_j+\delta_{jk} e_i \quad
      \text{for}\quad (i,j,k\leq 3)
      \quad \text{and}\quad (i,j,k\geq 4)\\
      [M_{ij}, e^*_k]&= -\delta_{ik} e^*_j+\delta_{jk} e^*_i \quad
      \text{for}\quad (i,j,k\leq 3) \quad \text{and}\quad (i,j,k\geq 4)
      ~.
    \end{aligned}
  \end{equation}
\end{footnotesize}
To continue with the computation, we introduce odd generators $Q_\pm$
which generate shifts along the the constant spinor $\psi_\pm$
parameterising the Killing spinors in \eqref{eq:killingspinors}.
As we have argued, the spinorial Lie derivative preserves the
space of Killing spinors.  Let $\xi$ be a Killing vector field.
Acting on a generic Killing spinor $\varepsilon(\psi_+, \psi_-)$,
$L_\chi$ will give a Killing spinor with different parameters
$\varepsilon( S^-_\xi\psi_+, S^+_\xi\psi_-)$.  This defines an action
of the Lie algebra of isometries on the space of Killing spinors,
whose structure constants are given by the constant matrices $S^-_\xi$
and $S^+_\xi$.  An explicit computation reveals that
\begin{equation}
  \label{eq:S0S1}
  \begin{aligned}[m]
    [e_+, Q_\pm]&=0\qquad [e_-, Q_+]=-\tfrac{\mu}{4} I Q_+\qquad
    [e_-, Q_-]=-\tfrac{\mu}{12}I Q_-\\
    [e_i, Q_+]&=-\tfrac{\mu}{6} I \Gamma_i \Gamma_+ Q_-\quad (i\leq3)\\
    [e_i, Q_+]&=-\tfrac{\mu}{12} I \Gamma_i \Gamma_+ Q_-\quad (i\geq4)\\
    [e^*_i, Q_+]&=-\tfrac{\mu^2}{18} \Gamma_i \Gamma_+ Q_-\quad (i\leq3)\\
    [e^*_i, Q_+]&=-\tfrac{\mu^2}{72} \Gamma_i \Gamma_+ Q_-\quad (i\geq4)\\
    [M_{ij}, Q_\pm]&=\tfrac{1}{2} \Gamma_{ij} Q_\pm \quad
    (i,j\leq3)~\text{and}~ (i,j\geq4)
  \end{aligned}
\end{equation}
Using the Clifford algebra, it is easy to check that
this forms a representation of $\fg_A \rtimes \fs_A$.

Finally we compute the bracket $\{\cdot,\cdot\}$ of odd generators of
the symmetry superalgebra.  For this let
$\varepsilon_1=\varepsilon(\psi_+,\psi_-)$ and
$\varepsilon_2=\varepsilon(\rho_+,\rho_-)$ be Killing spinors
associated with constant spinors $(\psi_+,\psi_-)$ and
$(\rho_+,\rho_-)$ as in \eqref{eq:killingspinors}, respectively.  Then
the vector $V=\bar\varepsilon_1 \Gamma^M \varepsilon_2\partial_M$ is a
linear combination of Killing vectors with constant coefficients:
\begin{multline}
  \label{eq:spinorsquare}
  V = - (\bar\psi_- \Gamma_+ \rho_-)\, \xi_{e_-} - (\bar\psi_+
  \Gamma_- \rho_+)\, \xi_{e_+} + \tfrac{\mu}{6}\sum_{i,j\leq 3}
  (\bar\psi_-\Gamma_+ I \Gamma_{ij} \rho_-)\, \xi_{M_{ij}} \\
  -\tfrac{\mu}{12} \sum_{i,j\geq 4} (\bar\psi_- \Gamma_+ I \Gamma_{ij}
  \rho_-)\, \xi_{M_{ij}} - \sum_{i\leq 3} \left( \bar\psi_+ \Gamma_i
  \rho_- +  \bar\psi_- \Gamma_i \rho_+ \right) \xi_{e_i}\\
    +\tfrac{3}{\mu} \sum_{i\leq 3} \left( \bar\psi_+ I \Gamma_i \rho_- +
    \bar\psi_-I \Gamma_i \rho_+ \right) \xi_{e^*_i} -
  \sum_{i\geq 4} \left( \bar\psi_+ \Gamma_i \rho_- + \bar\psi_-
    \Gamma_i  \rho_+ \right) \xi_{e_i}\\
  + \tfrac{6}{\mu} \sum_{i\geq 4} \left( \bar\psi_+ I\Gamma_i \rho_- -
    \bar\psi_- I \Gamma_i \rho_+ \right) \xi_{e^*_i}~,
\end{multline}
where $M_{ij}$ are the generators of the isotropy algebra $\fs_A$ of
$A$.  From this one can read the anticommutators of the odd generators
of the superalgebra as
\begin{equation}
  \label{eq:anticommKG}
  \begin{aligned}[m]
    \{Q_+, Q_+\}&=- \Gamma_-C^{-1} e_+ \\
    \{Q_+, Q_-\}&=- \sum^9_{i=1} \Gamma^iC^{-1} e_i+ \tfrac{3}{\mu}
    \sum_{i\leq 3} I\Gamma^iC^{-1} e^*_i +\tfrac{6}{\mu} \sum_{i\geq 4}
    I\Gamma^iC^{-1} e^*_i \\  
    \{Q_-, Q_-\}&=-\Gamma_+C^{-1} e_-+ \tfrac{\mu}{6}
    \sum_{i,j\leq3}\Gamma_+ I\Gamma^{ij}C^{-1} M_{ij}\\
    & \qquad {} - \tfrac{\mu}{12} \sum_{i,j\geq 4}
    \Gamma_+I\Gamma^{ij}C^{-1} M_{ij}~.
  \end{aligned}
\end{equation}
To summarise the (anti)commutation relations of the symmetry
superalgebra of the KG solution are given  by the
 equations \eqref{eq:S0S0}, \eqref{eq:S0S1} and
\eqref{eq:anticommKG}.

The symmetry superalgebra of an M-theory Hpp-wave solution
\eqref{eq:cwsoll} based on a generic CW space is the following.  The
commutators of the bosonic generators are given in \eqref{eq:liealg}.
This solution preserves only half of the supersymmetry and since the
Killing spinor $\varepsilon$ satisfies $\Gamma_+\varepsilon=0$, the
associated superalgebra contains only $Q_+$ generators.  It follows
that the remaining non-vanishing commutators and anticommutators are
\begin{equation}
  \label{eq:anticommHpp}
  \begin{aligned}[m]
    [e_-, Q_+]&=-\tfrac1{24} \Theta_{ijk} \Gamma^{ijk} Q_+\qquad 
    [M_{ij}, Q_+]= \half \Gamma_{ij} Q_+ \\
    \{Q_+, Q_+\}&=-\Gamma_-C^{-1} e_+~,
  \end{aligned}
\end{equation}
where $M_{ij}$ are generators of $\fs_A\cap\fs_\Theta$.  The symmetry
enhancement at the special point is dramatically illustrated by
comparing the symmetry superalgebra of the generic Hpp-wave, given by
\eqref{eq:liealg} and \eqref{eq:anticommHpp}, with that of the KG
solution, given by equations \eqref{eq:S0S0}, \eqref{eq:S0S1} and
\eqref{eq:anticommKG}.

\section{H-Branes}
\label{sec:hbranes}

Reducing the M-theory Hpp-wave found above and using U-duality, we can
construct H$p$-brane solutions in all type II supergravities.  The
reduction and U-duality for the metric and form-field strengths are
straightforward to perform.  We should only ensure that at every stage
we choose a CW space which is invariant under the direction that the
reduction and U-duality are performed.  In particular, this implies
that H$p$-branes for $p>0$ are associated with decomposable CW spaces.
 
The preservation of supersymmetry under reduction and U-duality is not
obvious.  This is because the Killing spinors depend nontrivially on
most of the eleven-dimensional coordinates and in particular $x^-$.
We shall come to this point later.
 
First we shall reduce the M-theory Hpp-wave to IIA along the direction
$y=\frac{1}{\sqrt{2}}(x^+ + x^-)$, where we have chosen $x^\pm=(\pm t
+ x^{10})/\sqrt{2}$.  We shall focus in the reduction of the Hpp-wave
given in \eqref{eq:cwsol}.  This is because the four-form field
strength vanishes along the direction $x^i$ for $i>3$ and the Killing
spinors depend trigonometrically on $x^-$.  Using the Kaluza--Klein
Ansatz leading to the string frame in IIA theory, we find the
following H0-brane solution:
\begin{equation}
  \label{eq:hzero}
  \begin{aligned}
    ds^2&= -\Lambda^{-\half} dt^2+\Lambda^{\half} ds^2(\RR^9)\\
    F_4&=-\tfrac{\mu}{\sqrt{2}} dt\wedge dx^1\wedge dx^2\wedge dx^3
    \\
    H_3&=\tfrac{\mu}{\sqrt{2}} dx^1\wedge dx^2\wedge dx^3 \\
    F_2&= -dt\wedge d\Lambda^{-1} \\
    e^{2\phi}&= \Lambda^{\tfrac32}~,
  \end{aligned}
\end{equation}
where $\Lambda=1+ \half A_{ij} x^i x^j$, $F_2, F_4$ are the RR field
strengths and $H_3$ is the NSNS three form field-strength.  Observe
that near the origin $x^i=0$, the metric of the H0-brane approaches
Minkowski space, the dilaton is constant, the two-form field strength
vanishes, and both $H_3$ and $F_4$ are constant.  Away from the
origin, the dilaton, and so the string coupling, increases or
decreases depending on whether along those directions $A$ is positive-
or negative-definite.  Along the directions that $A$ is
negative-definite the dilaton will eventually become complex
signalling an instability in the theory.  Despite the complicated
behaviour of the solution in ten dimensions, the associated Hpp-wave
solution in eleven dimensions is smooth and preserves at least one
half of the supersymmetry.  For example, the Hpp-wave associated with
the KG background has negative-definite $A$ and so the dilaton becomes
complex for all $x$ such that
\begin{equation*}
  \tfrac{\mu^2}{9} \left(\sum_{i\leq 3} (x^i)^2 + \tfrac14 \sum_{i\geq
  4} (x^i)^2 \right) > 1~.
\end{equation*}
The associated Hpp-wave preserves all supersymmetry.

To construct other H$p$-branes for $p>0$, one performs T-duality along
the directions $x^{3+q}$ for $1\leq q\leq p$ on the H0-brane
\eqref{eq:hzero}.  We now take $A_{ij} x^i x^j$ to be independent from
$x^{3+q}$ in which case the CW space becomes degenerate.  For this one
can use the T-duality rules of \cite{BHO}.  Although the application
of these rules is straightforward, the expressions for the H$p$-branes
are rather involved mainly due to the fact that the H0-brane has
non-vanishing form field-strengths from both the NSNS and RR sectors.
One can easily see that the metric and the dilaton for an H$p$-brane
are given by
\begin{equation*}
  \begin{aligned}[m]
    ds^2 &= \Lambda^{-\half} \left( -dt^2 + \sum^p_{q=1} (dx^{3+q})^2
    \right)\\
    & \qquad {} + \Lambda^{\half} \bigl( (dx^1)^2 + (dx^2)^2 +
    (dx^3)^2 + \sum_{i>p+3}^9 (dx^i)^2\bigr) \\
    e^{2\phi} &= \Lambda^{\tfrac{3-p}{2}}~.
  \end{aligned}
\end{equation*}

\section{Supersymmetry and H0-branes}
\label{sec:susiesH0}

To investigate whether supersymmetry is preserved in the construction
of H0-brane from the Hpp-wave above, we observe that the metric, the
four-form and Killing spinors are independent of $x^+$ and depend
trigonometrically on $x^-$.  For the case of M-theory Hpp-waves
\eqref{eq:cwsol} that preserve one half of the supersymmetry, we can
identify $x^+ \mapsto x^+ + 2\pi m \ell$ for any $\ell$ and $x^-
\mapsto x^- + \tfrac{4\pi}{\mu} n $, for some $m,n \in \ZZ$.  For even
$n$ we must take the trivial spin structure on this torus, whereas for
$n$ odd we must take the nontrivial spin structure on the $x^-$
circle, as the Killing spinors change sign.  To reduce the metric and
the four-form field strength of the Hpp-wave as above, we compactify
along the direction $x^{10}$ by setting $x^{10}\mapsto x^{10}+2\pi n
R$ for some $n\in \ZZ$ and $R\in \RR$; $x^{10}= \tfrac{1}{\sqrt{2}}
(x^++x^-)$.  Such identification is consistent with supersymmetry in
eleven dimensions provided that either $R=\sqrt{2} \ell=
\frac{2\sqrt{2}}{\mu}$ or $R=\sqrt{2} \ell=\frac{4\sqrt{2}}{\mu}$ and
an appropriate spin structure is chosen.

The above relations between the radius of compactification and the
periodicity of Killing spinors, although necessary, are not sufficient
for the reduced solution to preserve some of the supersymmetry of IIA
supergravity.  In fact, additional conditions are required for this to
be the case.  One such local condition is that the Lie derivative of
the Killing spinors should vanish along the Killing vector field
generated by translations along the compact direction $y$.  This is
similar to the condition required for solutions to preserve
supersymmetry after T-duality in the context of toric hyperkähler
manifolds \cite{GGPT}; T-duality in the context of supergravity can be
seen as hidden symmetry of the reduced theory.

For the case of the Hpp-wave, we can immediately see using
\eqref{eq:hppspinora} or \eqref{eq:hppspinorb} that
\begin{equation*}
  L_\xi \varepsilon\not=0
\end{equation*}
where $\varepsilon$ is the Killing spinor and $\xi\partial_y$ is the
vector field tangent to the compact direction; unless $\mu=0$ and the
Hpp-wave solution becomes Minkowski space.  Therefore we conclude that
the H0-brane does not preserve any supersymmetry of IIA supergravity.
This can be verified directly: for example, one sees that the dilatino
equation is not satisfied.  This is another example of reduction
breaking the supersymmetry of a solution in addition to those found in
\cite{Bakas, BKODuality, LPTAdS, DLP}.

The above conditions for the supersymmetries preserved by a solution
after reduction in the context of supergravity can be stated in terms
of the (anti)commutation relations of the symmetry superalgebra of the
original solution.  To see this, recall that the commutators of the
bosonic and fermionic generators of the symmetry superalgebra of a
solution are computed by evaluating the Lie derivative of the Killing
spinors along the Killing vectors associated with the bosonic
generators.  It is clear now that the unbroken supersymmetries of the
reduced solution are those fermionic symmetry generators of the
original solution that \emph{commute} with the bosonic generator that
is associated with translations along the compact direction.  In the
case of the Hpp-wave superalgebra \eqref{eq:anticommHpp}, since $e_-$
and $Q_+$ do not commute, the H0-brane does not preserve any
supersymmetry.

A similar analysis can be done for the case of Hpp-waves that preserve
all thirty-two supersymmetries in eleven dimensions.  The only
difference is that the identification in this case is $x^+ \mapsto x^+
+ 2\pi m \ell$ for any $\ell$ and $x^- \mapsto x^- +
\tfrac{12\pi}{\mu} n $, for some $m,n \in \ZZ$.  For even $n$ we must
take the trivial spin structure on this torus, whereas for $n$ odd we
must take the nontrivial spin structure on the $x^-$ circle.  Note
that if instead we identify $x^+ \mapsto x^+ + 2\pi m \ell$ for any
$\ell$, $x^- \mapsto x^- + \tfrac{4\pi}{\mu} n $, for some $m,n \in
\ZZ$ and take an appropriate spin structure, then only sixteen
supersymmetries are preserved in eleven dimensions.

The above analysis of supersymmetry suggests that if one does string
perturbation theory in the H0-brane background, the resulting spectrum
will not be supersymmetric and it will most likely contain tachyonic
modes.  However if additional possibly non-perturbative states are
included in the spectrum, then supersymmetry is restored in the
theory.

\section{HD-Branes and HNS-Branes}
\label{sec:HDHNS}

The solutions of M-theory that we have described in
Section~\ref{sec:HppCW} are suitable for describing the asymptotic
region of D0-branes in constant four-form field-strength.  Since a
constant four-form field-strength does not vanish at infinity, such
D0-brane solutions do not approach Minkowski spacetime at infinity.
Instead, as we shall see, they approach the solutions we have found in
Section~\ref{sec:HppCW}.  From the perspective of M-theory, the
supergravity description of D0-branes is as pp-waves.  It is then easy
to see that the solution that describes D0-branes in constant
four-form field strength is
\begin{equation}
  \label{eq:ppcwsol}
  \begin{aligned}[m]
    ds^2 &= 2 dx^+ dx^- + K
    (dx^-)^2 +  \sum_i dx^i dx^i\\
    F&= \mu\, dx^- \wedge dx^1 \wedge dx^2 \wedge dx^3~,
  \end{aligned}
\end{equation}
where $\mu$ is a constant and the field equations imply that
\begin{equation}
  \label{eq:ppcon}
  \delta^{ij}\partial_i\partial_j  K = - \mu^2~.
\end{equation}
A solution is
\begin{equation*}
  K = A_{ij} x^i x^j + \sum^N_{A=1} \frac{P_A}{|x-x_A|^7}~,
\end{equation*}
where $x_A$ are the positions of the D0-branes and $P_A$ are their
tensions.  It is straightforward to see that the above solution
preserves one half of the supersymmetry.  The Killing spinors and
symmetry superalgebra are those of the solutions of
Section~\ref{sec:HppCW} preserving one half of the supersymmetry.
Away from the D0-branes the solution approaches the solutions we have
described in Section~\ref{sec:HppCW}.  Near the centres, the solution
exhibits the same singularity structure as that of a pp-wave.

As in the case of H$p$-branes explained in the previous section to
compactify along a space like direction to IIA theory, we choose
$x^{10}= \frac{1}{\sqrt{2}} (x^+ + x^-)$.  In the string frame, the
ten-dimensional solution is given by
\begin{equation*}
  \begin{aligned}[m]
    ds^2&= -\Lambda^{-\half} dt^2+\Lambda^{\half} ds^2(\RR^9)\\
    F_4&=-\tfrac{\mu}{\sqrt{2}} dt\wedge dx^1\wedge dx^2\wedge dx^3
    \\
    H_3&=\tfrac{\mu}{\sqrt{2}} dx^1\wedge dx^2\wedge dx^3
    \\
    F_2&= -dt\wedge d\Lambda^{-1}
    \\
    e^{2\phi}&= \Lambda^{\tfrac32}~,
  \end{aligned}
\end{equation*}
where 
\begin{equation}
  \label{eq:lamd}
  \Lambda=1+ \half K~.
\end{equation}
If the matrix $A$ is positive-semidefinite, then the dilaton is large
at infinity and the string coupling constant is large.  On the other
hand if $A$ is not positive-semidefinite, then there are regions of
spacetime that the dilaton is not defined; that is, it becomes complex
and multivalued.  Nevertheless the existence of HD0-brane solutions in
supergravity theory suggests that there are sectors in the Hilbert
space of string theory in backgrounds with homogeneous fluxes which
are associated with D0-branes.  As in the case of H0-branes,
HD0-branes are not supersymmetric in the context of IIA supergravity.

One can construct HD$p$-branes for $p\leq6$ by T-dualising along the
directions $i>3$.  The solutions are similar to those presented in the
previous section for the associated H$p$-branes.  The only difference
is that the function $\Lambda$ appearing in the HD$p$-branes is given
in \eqref{eq:lamd}.  Using S-duality one can construct HNS-branes and
HM-branes.  The singularity structure of the new solutions is
different from that of the associated standard brane solution.  For
example consider the metric of a HNS5-brane in the string frame
located at $x=0$
\begin{equation*}
  ds^2 = ds^2(\RR^{1,5}) + \Lambda ds^2(\RR^4)~,
\end{equation*}
where $\Lambda=1+A_{ij} x^i x^j+\frac{P}{\delta_{ij} x^i x^j}$;
$i,j=1,\dots,4$.  For simplicity, we take $A_{ij}=\delta_{ij}$. The
HNS5-brane spacetime is complete with two asymptotic regions. The
near horizon geometry $|x|\rightarrow 0$ is $\RR^{1,6}\times S^3$ as
expected.  However the asymptotic geometry as $|x|\rightarrow \infty$
is not flat as the NS5-brane but $\RR^{1,5}\times C(S^3)$, where
$C(S^3)$ is a conformally flat cone over $S^3$.

\section{HF-Branes}
\label{sec:HF}

A large class of flux-branes, for example those associated with
certain Melvin solutions, can be constructed by reducing a Minkowski
spacetime \cite{DGGH1,DGGH2}.  This class of solutions leads to string
theory models which can be solved exactly \cite{RTconstant,RT}.  Such
backgrounds have been used to investigate tachyon instability in
closed string theory \cite{RT, GSflux, CG} and establish a conjecture
for the equivalence of type IIA and type OA string theory
\cite{BGtype0,CG}; for more recent progress see \cite{CHC}.  Instead
of using the Minkowski spacetime, one can reduce one of our Hpp-wave
solutions and in this way obtain homogeneous flux-brane solutions in
ten dimensions, or \emph{HF-branes} for short.

Instead of considering all configurations that can be found in this
way, we shall focus here in an HF-brane that arises from the
superposition of a Hpp-wave with the F7-brane.  To do this we consider
an Hpp-wave \eqref{eq:cwsoll} for which the isotropy group of $A$
contains an $\SO(2)$ subgroup and is invariant under translations
along an additional direction, say $x^3$.  In this case, it can always
be arranged such that
\begin{equation*}
  A_{ij} x^i x^j= b \left( (x^1)^2+(x^2)^2\right)+ \tilde A(x)
\end{equation*}
where $\tilde A(x) = A_{mn} x^m x^n$ for $m,n\geq 4$, whence
$\partial_3\tilde A=0$.

We perform the reduction along the orbits of the Killing vector field
$\xi=\partial_3+B R \partial_\varphi$, where $x^1=r \cos \varphi$,
$x^2=r\sin\varphi$, and where we have periodically identified $x^3\sim
x^3+2\pi R$.  Next we introduce the coordinate $\tilde
\varphi=\varphi- BR x^3$ which is constant along the orbits of $\xi$.
Expressing the Hpp-wave solution in terms of $\tilde \varphi$ and
$x^3$ coordinates, we perform the Kaluza--Klein reduction along $x^3$.
In particular, we shall consider three different choices for the
three-form $\Theta$ in $F$ of Hpp-wave \eqref{eq:cwsoll}.  First we
take $\Theta$ to be non-vanishing along the directions $x^1, x^2$.  In
this case, the Hpp-wave solution that we shall reduce to IIA is
\begin{equation*}
  \begin{aligned}[m]
    ds^2 &= 2 dx^+dx^- + \left(b \left((x^1)^2+(x^2)^2\right)+ \tilde
      A(x)\right) (dx^-)^2\\
    & {} \qquad + (dx^1)^2+(dx^2)^2+ \sum_{i>2} (dx^i)^2 \\
    F_4 &= dx^-\wedge dx^1\wedge dx^2\wedge \left(\sum_{i>3}u_i dx^i
    \right)
  \end{aligned}
\end{equation*}
where $b$ and $u_i$ are constants and $2 b+ \tr \tilde A = -\half u_m
u^m$.  Reducing along $x^3$, we find
\begin{equation}
  \label{eq:fhpp}
  \begin{aligned}[m]
    ds^2&= K^{\half} \biggl[2 dx^+dx^-+(b r^2+ \tilde A(x))
    (dx^-)^2 \\
    & {} \qquad + \sum_{i>3} (dx^i)^2+dr^2 \biggr] + K^{-\half}
    r^2 d\tilde\varphi^2 \\
    F_2&=d\left(\tfrac{Br^2}{K}\right) \wedge d\tilde \varphi \\
    H_3&=-BR r dx^-\wedge dr\wedge \left(\sum_{i>3}u_i dx^i\right)\\
    F_4 &= r dx^-\wedge dr \wedge d\tilde\varphi\wedge
    \left(\sum_{i>3}u_i dx^i\right)\\
    e^{\frac43 \phi}&=K~,
  \end{aligned}
\end{equation}
where $K=1+B^2 R^2 r^2$.  It is easy to see if one takes $A=0$, the
ten-dimensional solution becomes the F7-brane.  Observe though that
the worldvolume of the F7-brane in the presence of an Hpp-wave, is no
longer a Minkowski space.  This is an example of a brane whose
worldvolume is curved \cite{PRT}.  It is also straightforward to
show that the integral of $F_2$ over the plane with coordinates $(r,
\tilde\varphi)$ is finite as expected for an F-brane.

Two more cases arise by choosing $\Theta$ in different ways from the
case we have investigated above.  For example, if $F_4$ vanishes along
all directions $x^1, x^2, x^3$, then the IIA three-form vanishes and
the IIA four-form coincides with that of the Hpp-wave
\eqref{eq:cwsoll}.  Alternatively, one can choose $F_4$ to vanish
along the $x^1, x^2$ directions but to be non-vanishing along the
$x^3$ direction.  In this case the IIA four-form vanishes but
$H_3=i_3F_4$.  In both the above cases the metric and two-form field
strength of the IIA theory are as in \eqref{eq:fhpp}.

Since the Killing spinors of the generic Hpp-wave solution
\eqref{eq:cwsoll} do not depend on the coordinates $x^1, x^2, x^3$
involved in the compactification, the HF-brane that we have found in
\eqref{eq:fhpp} may preserve some of supersymmetry of IIA
supergravity.  The Lie derivative of the Killing spinors along the
compact direction vanishes unlike the case of H0-brane described in
Section~\ref{sec:hbranes}.  However now there is a global condition
depending on whether the Killing spinors are periodic or antiperiodic
with respect to the compact coordinate.  In the latter case
supersymmetry is broken.  The analysis is similar to that for the
standard flux-branes in \cite{CG}.

There are many other cases to be considered.  For example, one can
also reduce along a diagonal direction which involves the spacelike
direction of $x^-$ and $\varphi$.  In such a case, the IIA
configuration could be interpreted as a superposition of an H0-brane
with an F7-brane.  Many more solutions can be found by applying
U-duality to the above configurations.

\section{Concluding Remarks}
\label{sec:conc}

It is natural to ask whether there are more maximally supersymmetric
solutions of eleven-dimensional supergravity.  In fact there are not.
This result appears in \cite{KG} and confirmed in \cite{FOPMax}.
Therefore the moduli space of simply connected maximally 
supersymmetric solutions of
M-theory consists of a real line and an additional point.  The
coordinate on the real line can be chosen to be the scalar curvature
of the four-dimensional subspace of the maximally supersymmetric
$\AdS_4\times S^7$ and $\AdS_7\times S^4$ solutions.  The point with
vanishing scalar curvature can be identified with the Minkowski
vacuum.  The point corresponds to the KG solution, which does not have
free parameters.  Since the KG solution has vanishing scalar
curvature, it could also be identified with the point of vanishing
scalar curvature but in this case the moduli space would not be
Hausdorff.

The presence of ``unexpected'' maximally supersymmetric solutions in
eleven-dimensional supergravity raises the question whether similar
solutions exist in other supergravities.  Due to the
large number of symmetries, it is expected that string theory on such
backgrounds is solvable or amenable to a CFT/AdS type of conjecture.

\section*{Acknowledgements}

We thank Neil Lambert and Eric Bergshoeff for discussions, and Kelly
Stelle and Michael Duff for correspondence.  JMF is a member of EDGE,
Research Training Network HPRN-CT-2000-00101, supported by The
European Human Potential Programme.  GP is supported by a University
Research Fellowship from the Royal Society.  This work is partially
supported by SPG grant PPA/G/S/1998/00613.

%
%\bibliographystyle{amsplain}
%\bibliography{AdS,ESYM,Sugra,Geometry,CaliGeo}
%

\providecommand{\bysame}{\leavevmode\hbox to3em{\hrulefill}\thinspace}

\end{document}